\documentclass[usenatbib]{mn2e} \usepackage{psfig}\usepackage{graphicx}\usepackage{amsmath}\usepackage{amssymb}

\title[How extreme are the Wolf-Rayet Clusters in NGC\,3125?] {How extreme are the
Wolf-Rayet Clusters in NGC\,3125?\thanks{Based on
observations made with ESO telescopes at the Paranal Observatory under
programme ID 074.B-0108 and with archival ESO VLT and NASA/ESA Hubble
Space Telescope data, obtained from the ESO/ST-ECF Science Archive
Facility.}}

\author[L.\,J.\,Hadfield et al.] {L.\,J.\ Hadfield,\thanks{E-mail:
  l.hadfield@shef.ac.uk} P.\,A.\ Crowther\\ Department of Physics and Astronomy, University of
Sheffield, Sheffield, S3 7RH, UK\\}

\date{}

\begin{document}
  \maketitle
  
\begin{abstract}  
We reinvestigate the massive stellar content of the irregular dwarf
galaxy NGC\,3125 (Tol~3) using VLT/FORS1 imaging and spectroscopy,
plus archival VLT/ISAAC, HST/FOC and HST/STIS datasets. FORS1
narrow-band imaging confirms that the NGC\,3125 A and B knots
represent the primary sites of Wolf-Rayet (WR) stars, whilst HST
imaging reveals that both regions host two clusters. Both clusters
within region A host WR stars (A1 and A2), for which the optically
fainter cluster A2 is heavily reddened. It is not clear which cluster
within region B hosts WR stars. Nebular properties are in good
agreement with previous studies and infer a LMC-like metallicity of
log(O/H)+12$\sim$8.3. LMC template mid-type WN and early-type WC
spectra are matched to the observed blue and red WR bumps of A1 and B,
permitting the contribution of WC stars to the blue bump to be
quantified. From our FORS1 spectroscopy we obtain N(WN5--6:WC4)=105:20,
$\sim$55:0, 40:20 for clusters A1, A2 and B1\,+\,2, respectively.  Our
results are a factor of $\sim$3 lower than previously reported by
optical studies as a result of a lower H$\alpha$/H$\beta$ derived
interstellar reddening. Using Starburst99 theoretical energy
distributions to estimate O star populations for each cluster, we find
N(WR)/N(O)=0.2 for A1 and 0.1 for A2 and the clusters within region B.
From H$\alpha$ narrow-band imaging, the O star content of the Giant
H\,{\sc ii} regions A and B is found to be a factor of 5--10 times
higher than that derived spectroscopically for the UV/optically bright
clusters, suggesting that NGC\,3125 hosts optically obscured young
massive clusters, further supported by VLT/ISAAC K band imaging.
Archival HST/STIS UV spectroscopy confirms the low interstellar
reddening towards A1, for which we have determined an SMC
extinction law for NGC\,3125, in preference to an LMC or starburst
law. We obtain N(WN5--6)=110 from the slit loss corrected
He\,{\sc ii} $\lambda 1640$ line flux.  This is in excellent agreement
with optical results, although it is a factor of 35 times lower than
that inferred from the same dataset by \citet{chandar04}.  The
discrepancy is due to an anomalously high interstellar reddening
derived from their use of the generic starburst extinction law. Highly
discrepant stellar populations may result in spatially resolved star
forming regions from UV and optical studies through use of different
extinction laws.

\end{abstract}



\section{Introduction}
\label{introduction}

In order to understand galaxy formation and evolution we need to be
able to accurately map the star formation history of the universe \citep{madau96}.
Central to this topic are starburst galaxies, a class of object which
display characteristics associated with massive, violent bursts of
star formation.  In the local Universe only a handful
of starburst galaxies are responsible for a quarter of the entire
high-mass star formation \citep{heckman98}. We need to be
able to understand nearby starbursts if we are to interpret the
observations of distant star forming regions e.g. Lyman break galaxies
\citep[LBGs,][]{steidel96}.

A subset of these galaxies are called ``Wolf-Rayet galaxies'', since
their integrated spectra display the broad emission signatures
associated with Wolf-Rayet (WR) stars; the highly evolved descendants
of the most massive O stars. Spanning a wide variety of morphological
types, WR galaxies are observed in a wide variety of environments on a
local (D$<$100Mpc) scale and out to high redshift.  WR stars are
exclusively associated with young stellar populations ($\sim$5Myr), so
WR galaxies represent an excellent diagnostic of recent star
formation.

Recently, \citet{shapley03} demonstrated that the composite spectrum
of {\it z}\,$\sim$\,3 LBGs displayed broad He\,{\sc ii} $\lambda 1640$
emission consistent with the presence of WR stars.  The recent
ultraviolet HST/STIS spectral survey of local starburst galaxies
\citep{chandar04} revealed weak He\,{\sc ii} $\lambda 1640$ emission in
most cases. Of the 18 galaxies included in their survey, the super
star cluster NGC\,3125-1 \citep[alias NGC\,3125-A in][]{vc92} showed
the most prominent He\,{\sc ii} $\lambda 1640$ emission.




NGC\,3125 (Tol\,3) is a nearby \citep[D=11.5Mpc --][]{schaerer99} blue
compact dwarf galaxy.  Observations have shown that the galaxy is
dominated by a central starburst region which consists of two main
emission knots, NGC\,3125-A and -B.  From UV spectroscopy,
\citet{chandar04} estimate a WR population of $\sim 5000$ and
N(WR)/N(O)$\geq 1$ for NGC\,3125-A. The latter is completely
unexpected for the LMC-like metallicity of NGC\,3125.  In contrast,
optical studies of NGC\,3125-A infer a WR population of only $\sim
500$ and N(WR)/N(O)$\sim$ 0.1, an order of magnitude lower.  WR
populations in other nearby galaxies common to optical and UV surveys
are found to be consistent to within a factor of two. If NGC\,3125-A
is a local analogue for LBGs it is necessary to reconcile optical and
UV line techniques for this galaxy.

Here we reinvestigate the WR population of NGC\,3125 using new
VLT/FORS1 imaging and spectroscopy, supplemented by archival VLT/ISAAC imaging
and HST imaging and spectroscopy.  This paper is organised as follows:
VLT and HST observations of NGC\,3125 are discussed in
Section~\ref{obseverations}. Section~\ref{nebular} describes nebular
properties derived for the two WR clusters using optical diagnostics.
In Section~\ref{stellar}, WR and O star populations for each cluster
are estimated using VLT/FORS1 imaging and spectroscopy.  In addition,
the WR population of NGC\,3125-A is estimated from UV HST/STIS
spectroscopy. Finally, we draw our conclusions in Section~\ref{conclusions}.


\section{Observations and data reduction}
\label{obseverations}
We have observed NGC\,3125 with the ESO Very Large Telescope UT2
(Kueyen) and Focal Reduced/Low dispersion Spectrograph \#1 (FORS1), a
2048 $\times$ 2046 pixel Tektronix detector.  Observations were made
using the high resolution collimator which covers a 3.4\arcmin
$\times$ 3.4\arcmin\ field of view, with a plate scale of 0.1\arcsec
pixel$^{-1}$. Photometric observations of NGC\,3125 were acquired during
November 2004 and January 2005, with spectroscopic data following in
February 2005. Details of the observations, including the DIMM seeing
can be found in Table \ref{ngc3125obs}.

To supplement VLT/FORS1 observations we have retrieved VLT/ISAAC imaging and
HST imaging and spectroscopy of NGC\,3125 from the ESO/ST-ECF archive. 

\subsection{VLT Imaging}
\label{vlt:images}

FORS1 was used to obtain narrow-band images centred on $\lambda 4684$
and $\lambda 4781$ (FWHM=66\AA\ and 68\AA\ respectively).  The
$\lambda 4684$ filter coincides with the strong N\,{\sc iii} ($\lambda
4640$\AA), C\,{\sc iii} ($\lambda 4650$\AA), and He\,{\sc ii}
($\lambda 4686$\AA) WR emission lines, whereas the $\lambda 4781$
samples a wavelength region relatively free from emission, providing a
measure of the continuum level.  In addition, narrow-band on- and
off-H$\alpha$ images ($\lambda$ 6563,\,6665\AA, \,
\mbox{FWHM}=61,\,65\AA) were acquired along with broad-band B images.

\begin{table}
\caption{VLT/FORS1 observation log for NGC\,3125}
\label{ngc3125obs}
\begin{center}
\begin{tabular}{clll}
\hline
Date & Observation & Exposure & DIMM seeing\\
&&\multicolumn{1}{c}{(sec)} & \multicolumn{1}{c}{(\arcsec)}\\
\hline
\multicolumn{4}{c}{Imaging}\\
\hline
2004-Nov-08&  $\lambda$4684  & 60, 900 &0.42--0.43\\
&   $\lambda 4781$ & 60, 900 & 0.42,0.38\\
2004-Nov-17&  B & 10, 60, 600&0.58--0.59\\
& $\lambda$6665&10, 60, 600&0.45--0.57\\
\smallskip
2005-Jan-17&  $\lambda$6563 & 10, 60, 600& 0.67--0.79\\
\hline
\multicolumn{4}{c}{Spectroscopy}\\
\hline
2005-Jan-01&300V&4$\times$ 600&0.60\\
\hline
\end{tabular}
\end{center}
\end{table}  

\begin{figure*}
\begin{center}
\begin{tabular}{cc}
\includegraphics[width=7cm,clip, angle=0]{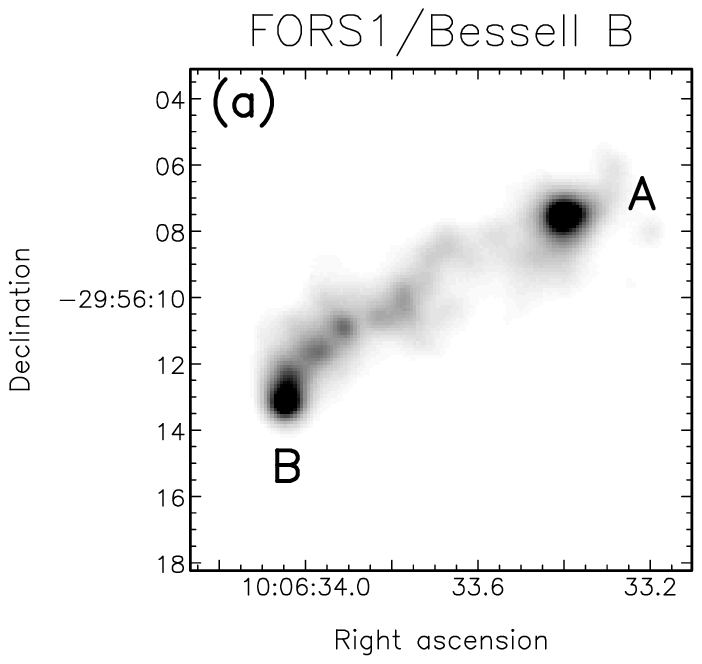}&
\includegraphics[width=7cm,clip, angle=0]{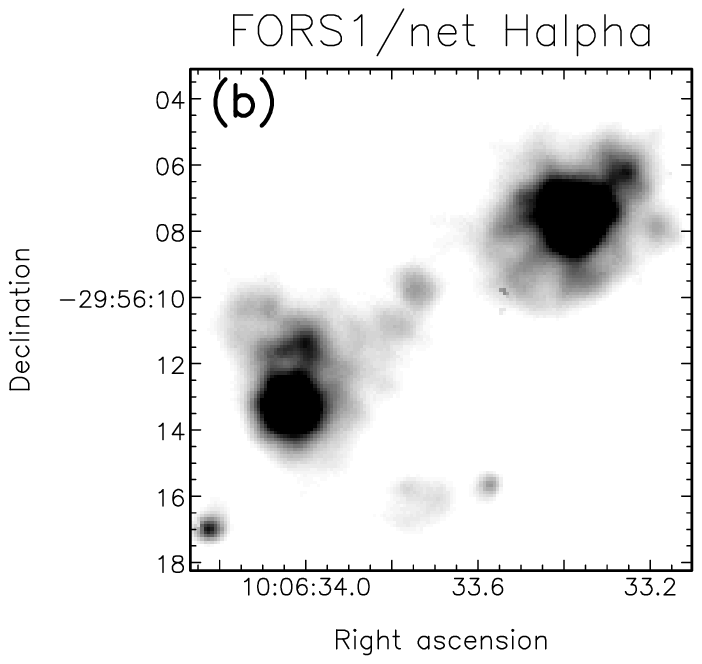}\\
\includegraphics[width=7cm,clip, angle=0]{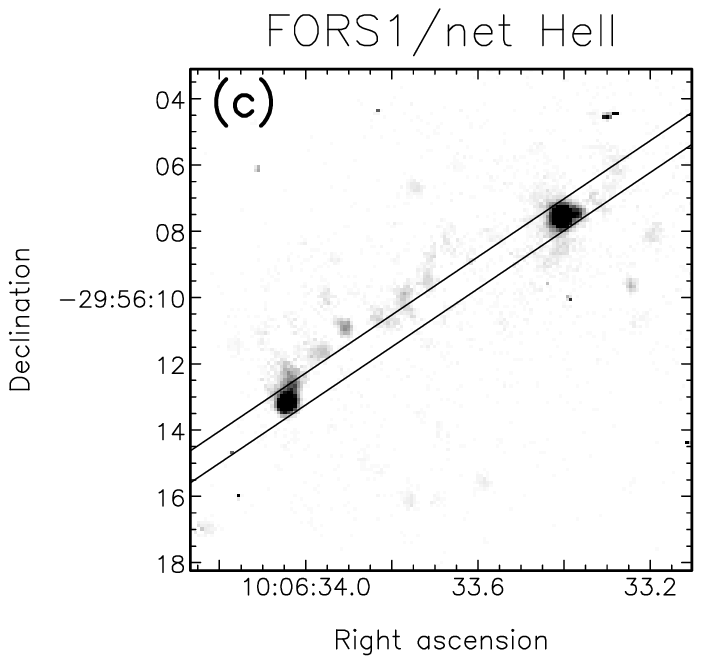}&
\includegraphics[width=7cm,clip, angle=0]{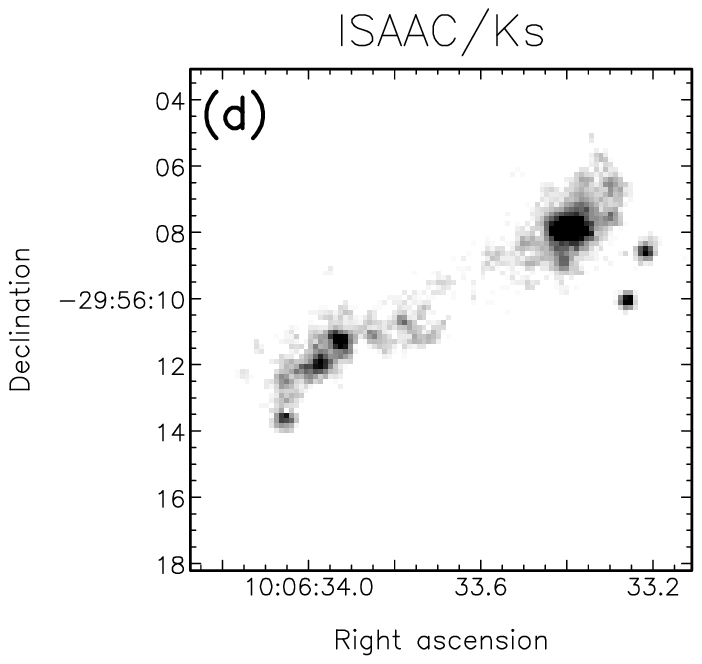}\\
\end{tabular}
\end{center}
\caption{15\arcsec $\times$ 15\arcsec\ VLT/FORS1 and VLT/ISAAC archival images of NGC\,3125. For our assumed distance 11.5Mpc, the physical region
  illustrated equates to 750 $\times$ 750\,pc.   a) Bessell B image showing the morphology of the central starburst.  Regions A and B have been
  marked along with the slit position of spectroscopic observations.
  b) High contrast, net H$\alpha$ image.  c) Difference between $\lambda 4684$ and $\lambda 4871$ filters, showing He\,{\sc ii}\,$\lambda 4686$ /
  C\,{\sc iii}\,$\lambda 4650$ emission. d) Archival VLT/ISAAC K$_{\mbox{s}}$ image
  of NGC\,3125. North is up and East is to the left on all images.}
\label{figure:images}
\end{figure*}

Images were reduced following standard reduction procedures (i.e
debiased, flat field corrected and cosmic ray cleaned) using {\sc
iraf} and {\sc starlink} packages.  We present continuum, net
H$\alpha$ images and net $\lambda$4684 images obtained with FORS1 in Figure
\ref{figure:images}.

The optical appearance of NGC\,3125 is that of an amorphous elliptical
\citep{schaerer99}, dominated by a bright central starburst.  In
Figure~\ref{figure:images}(a) we present a B band image of the central
starburst region taken in excellent seeing conditions
($\sim$0.6\arcsec).  The continuum emission is dominated by two
emission knots A and B \citep{vc92}. A corresponds to
the slightly brighter knot and is located $\sim$10\arcsec\ to the NW
of B. In the image, knot A appears to be partially resolved
into two components whereas knot B does not.  Several fainter
emission knots appear to link the two dominant regions.

Figure~\ref{figure:images}(b) shows the VLT/FORS1 net H$\alpha$ image
of the central starburst.  Nebular emission is concentrated on the two
principal knots with region A extending 300pc and region B 200pc.
This large size and H$\alpha$ luminosities of 3.2$\times 10^{40}$ and
2.0$\times 10^{40}$\,erg\,s$^{-1}$ for A and B, respectively (see
Section~\ref{halpha}), indicate that A and B are typical of
extragalactic Giant H\,{\sc ii} regions which host multiple stellar
clusters \citep{kennicutt84}.


We have searched for characteristic WR signatures NGC\,3125 by
subtracting the $\lambda 4871$ image from the $\lambda 4684$ image,
which is shown in Figure~\ref{figure:images}(c).  Regions A and B are
clearly identified as the primary hosts of WR stars. Region A contains
a fainter emission component $\sim$0.4\arcsec to the W of the main
source.  This fainter component contributes $\sim$15\% of the total
$\lambda$4684 emission.

Photometry of knots A and B was performed using the aperture
photometry package {\sc phot} within {\sc iraf}.  Instrumental zero
points were derived by observing LTT3864 ($B=12.7$) and GD108
($B=13.3$) spectrophotometric standard stars.  


Magnitudes for source A include the fainter component to the W of the
knot.  We find m$_{\lambda 4686}$=17.0$\pm 0.1$ and m$_{\lambda
4781}$=17.2$\pm$0.1. Photometric uncertainties have been estimated
from background variations, contamination by nearby emission sources
and the aperture size.  An extremely close emission knot to the NW of
source B made aperture photometry more difficult, reflected by larger
photometric uncertainties, for which we derive m$_{\lambda 4686}$
and m$_{\lambda 4781}$ magnitudes of 17.5$\pm$0.2 and 17.7$\pm$0.2
respectively.

To complement our imaging dataset we have retrieved archival VLT/ISAAC
images obtained on 18 April 2002 under excellent seeing conditions of
FWHM$\sim$0.4\arcsec.  These observations comprise broad-band K$_s$
imaging (5 on source plus 2 off-source exposures of 5\,s) taken using
the SW mode (Rockwell-Hawaii detector, plate-scale =
0.148\,arcsec\,pixel$^{-1}$). Images were calibrated relative to the
2MASS source 10063237-2956190 (K$_{\mbox{s}}$=15.0\,mag) which was
present in this 2.5\arcmin$\times$2.5\arcmin field.

In contrast to optical images, Figure~\ref{figure:images}(d) shows
that in the near-IR, knot A is much brighter than B, with
K$_{\mbox{s}}$=14.5 and 16.1\,mag, respectively.  In addition to the
optically dominant sources, there are two bright knots to the NW of B,
with a combined magnitude of K$_{\mbox{s}}$=15.0\,mag.  These are
again more prominent than in optical images, suggesting the presence
of partially obscured star forming regions.  We will return to this in
Section~\ref{halpha}.

\subsection{VLT Spectroscopy}

Spectroscopic observations of NGC\,3125 were undertaken on 1 February
2005 using FORS1 and the high resolution collimator in seeing
conditions of $\sim$0.6\arcsec.  Spectra were acquired using a
0.8\arcsec slit and 300V grism centred along the two main emission
knots (PA=-124$^{\circ}$; see Figure~\ref{figure:images}(c)).
The extracted spectra covered a wavelength range of 3300--8600\AA\ with a
dispersion of 2.6\AA\,pixel$^{-1}$ and resolution of $\sim$ 15\AA\ (as
measured from comparison arc lines).

Data were prepared and processed using standard techniques and {\sc
iraf} and {\sc starlink} packages i.e. bias subtracted, flat field
corrected, extracted and flux / wavelength calibrated.  Care was taken
during the extraction process to ensure neighbouring emission knots
did not contaminate the aperture or the background subtraction.  The
spectrophotometric standard Feige~66 was observed in order to
relatively flux calibrate the spectra.  

\begin{figure}
\centerline{\psfig{figure=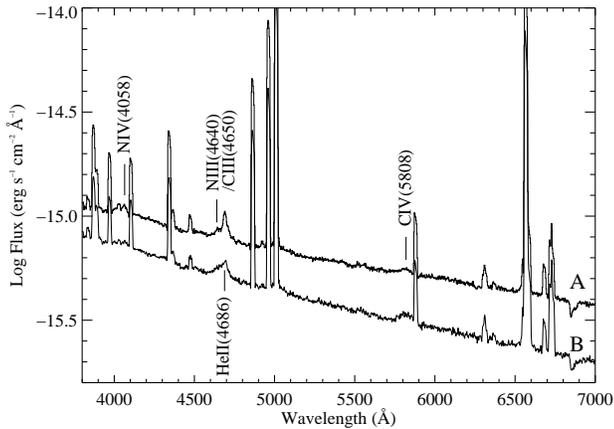,width=\columnwidth,angle=90.}}
\caption{VLT/FORS1 optical spectra of NGC\,3125-A and B.  Spectra have
  been velocity corrected \citep[V$_r$ = 865km\,s$^{-1}$][]{Lauberts} and WR
  emission features have been marked.}
\label{spectra}
\end{figure}

Absolute flux calibration was achieved by comparing synthetic-filter
photometry ($\lambda_c = 4684, FWHM=66$\AA) to $\lambda 4684$
photometry.  We derive a slit correction factor of 1.3$\pm$0.2 for knot A,
which dominates the accuracy of our final flux calibrated spectra.  An
identical correction factor is found for knot B.  We expect our
absolute flux calibration to be correct to $\sim$15\%. Extracted
spectra of NGC3125-A and B are shown in Figure \ref{spectra}, in which
the principal WR emission features are identified, including weak N\,{\sc
iv} $\lambda 4058$.

\subsection{HST Imaging}
\label{hst:images}
To compliment our FORS1 dataset we have retrieved archival HST images,
including a STIS/LONG\_PASS acquisition image (t = 40\,s, $\lambda_{c}$
= 7200\AA) from programme GO 9036 (P.I. C.~Leitherer) of region A and
a FOC/F220W\footnote{FOC/F220W image is pre-COSTAR.} image (t
=500\,s, $\lambda_{c}$ = 2280\AA) from programme GO 4800
(P.I. P.~Conti) of the central starburst region.

\begin{figure}
\begin{center}
\begin{tabular}{c}
\includegraphics[width=6cm, angle=0]{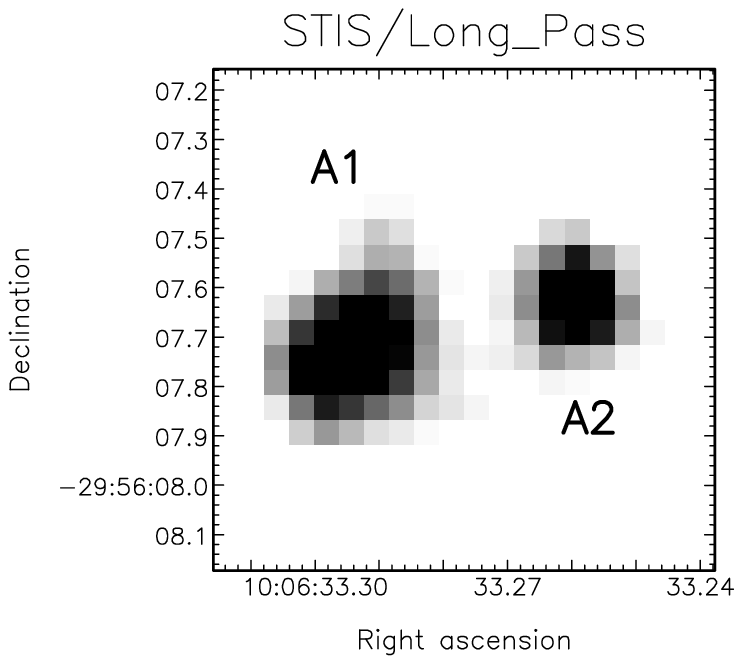}\\
\includegraphics[width=6cm, angle=0]{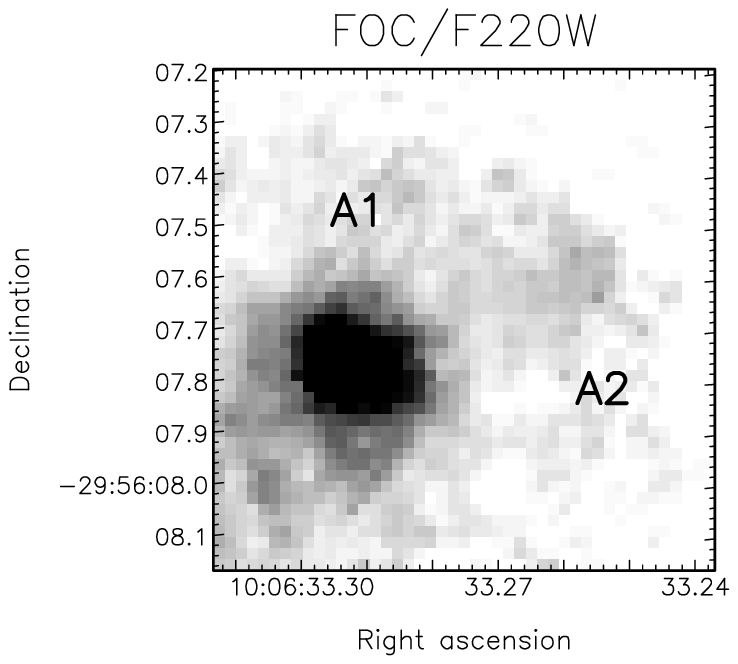}\\
\end{tabular}
\caption{1\arcsec $\times$ 1\arcsec\ (50$\times$50\,pc)
STIS/Long\_Pass acquisition (top) and pre-COSTAR HST/FOC F220W
(bottom) images of regions A.  The two clusters A1 and A2 are
marked.  North is up and East is to left.}
\label{HST:regionA}
\end{center}
\end{figure}

The STIS/LONG\_PASS acquisition image resolves region A into two
clusters, separated by $\sim0.5$\arcsec or 25pc at the distance of
NGC\,3125 (upper panel of Figure~\ref{HST:regionA}). We designate
these two components A1 and A2, with A1 corresponding to the brighter
of the pair to the east. The components have FWHM of 0.17\arcsec and
0.13\arcsec, and a flux ratio of
F$_{\tiny{\mbox{A1}}}$/F$_{\tiny{\mbox{A2}}} \sim$1.9 at 7200\AA,
although an absolute flux calibration was not possible.

In the HST/FOC F220W image A1 is bright whereas A2 is barely detected,
with
F$_{\tiny{\mbox{A1}}}$(F220W)/F$_{\tiny{\mbox{A2}}}$(F220W)\,$\sim$10
(lower panel of Figure~\ref{HST:regionA}).  Assuming that A1 and A2
possess comparable intrinsic energy distributions -- which is
reasonable since both clusters host WR stars
(Section~\ref{vlt:images}) -- A2 must suffer from significantly higher
extinction.  Aperture photometry reveals
F$_{\tiny{\mbox{A1}}}$(F220W)=(1.1$\pm 0.2) \times
10^{-15}$erg\,s$^{-1}$cm$^{-1}$ and
F$_{\tiny{\mbox{A2}}}$(F220W)\,$\leqslant$\,3$\times
10^{-16}$erg\,s$^{-1}$cm$^{-1}$.

\begin{figure}
\begin{center}
\includegraphics[width=6cm, angle=0]{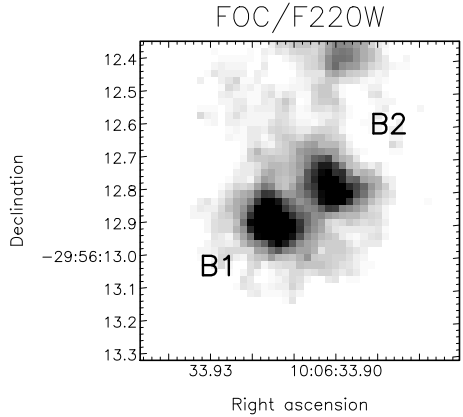}\\
\caption{1\arcsec $\times$ 1\arcsec\ (50 $\times$50\,pc) pre-COSTAR HST/FOC F220W
  ($\lambda_{c}= 2280$\AA) image of region B. North is up and East is to left.  }
\label{HST:regionB}
\end{center}
\end{figure}

The HST/FOC image also resolves region B into two components
(Figure~\ref{HST:regionB}) -- B1 and B2, with the brighter component,
B1 to the SE with
F$_{\tiny{\mbox{B1}}}$(F220W)/F$_{\tiny{\mbox{B2}}}$(F220W)$\sim$1.2.
It is not clear which cluster within region B hosts the WR
population since these are spatially unresolved in our VLT/FORS1
imaging. High spatial resolution optical imaging is not yet available.
It was not possible to perform aperture photometry on B1 and B2
individually in these pre-COSTAR images, we therefore derive a total
UV flux of $(1.6\pm0.2)\times 10^{-16}$erg\,s$^{-1}$cm$^{-1}$ for
region B.

\subsection{HST Spectroscopy}

Archival HST/STIS spectroscopy of region NGC\,3125-A1
\citep{chandar04} was obtained in programme GO 9036
(P.I.~C.~Leitherer) and comprised two exposures with the G140L grating
(FUV-MAMA detector), of duration 1050 and 2925\,s, plus one exposure
with the 230L grating (NUV-MAMA detector) using the
52$\times$0.2\arcsec slit.  The STIS MAMA detectors have a plate scale
of 0.024\arcsec\,pixel$^{-1}$.  Spectra were extracted over a 13 pixel
(0.3\arcsec) aperture and combined to provide a complete wavelength coverage
of 1175--3100\AA, with a spectral resolution of $\sim$3\,pixels (1.8\AA).

The UV spectrum of A1 has been presented by \citet{chandar04} who
noted the strong He\,{\sc ii} $\lambda$1640 emission feature, for
which we measure an equivalent width of 6.9$\pm$0.8\AA\ and
FWHM$\sim$4.8$\pm$0.7\AA.  Absolute flux calibration has been achieved
by comparing our HST/STIS spectra with the F220W flux of A1. We derive
spectral slit losses of $\sim$40$\pm30$\%.

\section{Nebular analysis}
\label{nebular}

In the following section we will derive the nebular properties of
NGC\,3125-A and B.  Observed (F$_{\lambda}$) and dereddened
(I$_{\lambda}$) nebular line fluxes with respect to H$\beta$=100 are
presented in Table~\ref{ngc3125:nebular}.

The nebular analysis was performed using the {\sc starlink} package
{\sc dipso} with line fluxes being determined using the ELF (emission
line fitting) routine.  The emission line profiles were non-gaussian
in the FORS1 spectra (comparison arc lines displayed the same profile,
apparently due to the use of the high-resolution collimator for spectroscopy)
and were modelled using template profiles created from the
data. [O\,{\sc iii}] and [O\,{\sc ii}] emission lines were modelled
using [O\,{\sc iii}] $\lambda 5007$ as a template profile,
H$\alpha,\beta,\gamma$ and [N\,{\sc ii}] were modelled using the H$\beta$
emission line as a template.

 
\begin{table}
\caption{Observed (F$_{\lambda}$) and dereddened (I$_{\lambda}$)
  nebula line fluxes of WR regions A and B within NGC\,3125.  Line
  ratios are normalised to H$\beta$=100.  We present observed and
  dereddened H$\beta$ fluxes in the final row (erg\,s$^{-1}$cm$^{-2}$).  Reddening corrections of E$_{B-V}$=0.24
  and 0.21 for A and B respectively, include a Galactic foreground
  reddening of 0.08\,mag.  }
\begin{center}
\begin{tabular}{ll@{\hspace{1mm}}l@{\hspace{1mm}}l@{\hspace{1mm}}l@{\hspace{1mm}}l}
\hline
\hline $\lambda_{\mbox{rest}}$&&\multicolumn{2}{c}{ --------- A ---------}&\multicolumn{2}{c}{----------- B ---------}\\
(\AA)&ID&F$_{\lambda}$&I$_{\lambda}$&F$_{\lambda}$&I$_{\lambda}$\\
\hline 3727&[O\,{\sc ii}]&83.5&99.0&109.2&123.9\\
4340&H$\gamma$&40.0&44.7&40.3&43.3\\ 
4363&[O\,{\sc iii}]&3.9&4.6&2.8&2.9\\ 
4861&H$\beta$&100&100&100&100\\ 
4959&[O {\sc iii}]&205.7&203.2&169.8&162.3\\ 
5007&[O\,{\sc iii}]&649.3&615.2&524.2&496.0\\
6563&H$\alpha$&371.4&282.9&351.2&279.2\\ 
6583&[N\,{\sc ii}]&8.7&6.6&11.5&9.2\\ 
7330&[O\,{\sc ii}]&3.7&2.7&3.6&2.7\\ 
4861&H$\beta$&6.95$\times 10^{-14}$&1.56$\times 10^{-13}$&3.83$\times 10^{-14}$&7.55$\times 10^{-14}$\\ 
\hline
\end{tabular}
\end{center}
\label{ngc3125:nebular}
\end{table}

\begin{table}
\begin{center}
\label{ngc3125:properties}
\caption{Summary of nebular properties of NGC\,3125 clusters A and B.}
\begin{tabular}{lll}
\hline\hline
&\multicolumn{1}{c}{A}&\multicolumn{1}{c}{B}\\
\hline
T$_{e}$(O\,{\sc ii})\,(K)&12\,100$\pm 1\,000$&10\,800$\pm 500$\\
T$_{e}$(O\,{\sc iii})\,(K)&10\,500$\pm 500$&9\,800$\pm 500$\\
n$_{e}$\,(cm$^{-3}$)&140$^{\dag}$&140$^{\dag}$\\
O$^{+}$/H &(1.8$\pm 0.2)\times 10^{-5}$ &(3.5$\pm 0.2) \times 10^{-5}$\\
O$^{2+}$/H &(1.9$\pm 0.1)\times 10^{-4}$ & (1.9$\pm 0.1)\times 10^{-4}$\\
$\log$(O/H) + 12&8.32$\pm 0.03$&8.35$\pm 0.03$\\
\hline
\end{tabular}
\end{center}
\small {$^{\dag}$ Electron densities were derived for [S\,{\sc ii}] ratios of VC92.}\\
\end{table}  

\subsection{Interstellar reddening}
\label{reddening}
Estimates of the interstellar reddening have been made using the
Balmer line ratios H$\alpha$:\,H$\beta$:\,H$\gamma$.  Nearby [N\,{\sc
ii}] emission has been accounted for when measuring observed H$\alpha$
fluxes.  Assuming Case B recombination theory for electron densities
of 10$^{2}\,\mbox{cm}^{-3}$ and a temperature of $10^4$\,K
\citep{hummer87} we deduce average total $E_{B-V}\ \mbox{values of}\
0.24\ \mbox{and}\ 0.21$ for A and B, respectively.

Foreground Galactic reddening ($E_{B-V}$) towards NGC\,3125 of $E_{B-V}$=0.08\,mag
\citep{schlegel98} was accounted for using a standard Galactic extinction law
\citep{seaton79}.  The point-like appearance of the two regions on ground-based images
suggest that a \citet{calzetti94} starburst obscuration law is inappropriate. We
therefore choose to use the \citet{bouchet85} SMC extinction law to deredden our spectra
(see Section~\ref{comparison:uv}).

Underlying stellar H$\alpha$ and H$\beta$ absorption from early-type
stars is estimated to be W$_{\lambda}$\,$\sim$2\AA.  In our spectra,
H$\alpha$ and H$\beta$ equivalent widths are found to be $\sim$550\AA\
and $\sim$100\AA\ for both regions.  Propagating this correction
through calculations leads to an uncertainty in E$_{B-V}$ of
$\pm$0.01\,mag.

From a comparison between the STIS/Long\_Pass acquisition and FOC/F220W
images (Figure ~\ref{HST:regionA}),
cluster A2 appears to suffer a much higher extinction than A1, with
F$_{\tiny{\mbox{A1}}}$/F$_{\tiny{\mbox{A2}}} \sim$1.9 at 7200\AA\ and
$\sim$10 at 2280\AA.  If we assume that A1 and A1 possess identical
intrinsic flux distributions, we estimate that
E$_{B-V}^{\tiny{\mbox{INT}}}$(A2)$\sim$0.5\,mag. 

\citet{kunth81} derived a total E$_{B-V}$=0.40 from
H$\beta$:H$\delta$:H$\gamma$ intensity ratios\footnote{\citet{kunth81}
quote a colour excess of E$_{B-V}$=0.28\,mag in their Table~1, the
origin of which is not given.  It is not clear which value has been
used in their subsequent analysis.}. Higher (and weaker) members of
the Balmer series can be significantly affected by underlying stellar
absorption, which was not accounted for in their analysis.

\citet[][hereafter VC92]{vc92} also studied the nebular properties of
NGC\,3125-A and B.  Using H$\alpha$:H$\beta$ line ratios they derived
internal reddenings of E$_{B-V}^{\tiny{\mbox{INT}}}$=0.40 and 0.64,
respectively, significantly higher than those obtained here.  A higher
extinction for region B than region A is inconsistent with our UV and
optical photometry, if we assume identical flux distributions for
these regions, since B is brighter in the UV whilst A is brighter
optically.

VC92 based their determination on composite blue ($\lambda
< 5400$\AA, photometric) and red ($\lambda >$4500\AA, non-photometric)
spectra obtained 12 months apart.  The red spectrum, which was scaled
to the blue continuum, resulted in a H$\beta$ flux
20\% larger than that measured in the blue spectrum.  Based upon their red
dataset alone, an extinction of NGC\,3125-A fully consistent with
the present result would have been obtained \citep[][private
communication]{vacca}.


\subsection{Electron temperature, density  \& oxygen abundance}

Electron temperatures, T$_e$, for regions A and B have been derived
from the temperature diagnostics [O\,{\sc ii}]\,3727/7325 and [O\,{\sc
iii}]\,(4959+5007)/4363 \AA\ for the line ratios presented in
Table~\ref{ngc3125:nebular}.  Electron temperatures were calculated
using the five-level atom calculator {\sc temden} within {\sc iraf}
for a constant electron density of 100\,cm$^{-3}$.

Errors on T$_{e}$[O\,{\sc ii}] were based on the 5\% measurement error estimated for the [O\,{\sc
    ii}] 7325\AA\,line, while T$_{e}$[O\,{\sc iii}] uncertainties
    were based on the 10\% formal error given for [O\,{\sc iii}] 4363\AA.

Electron temperatures derived here agree to within
the errors of those reported by VC92 who found 10\,200\,K for region A and 10\,300\,K for region B.

Due to the low spectral resolution of our data [S\,{\sc
 ii}]\,6716/6731\AA\ nebula lines are not resolved, preventing a
 direct estimate of the electron density $n_e$.  For completeness, we
 have estimated $n_e$ by combining our electron temperatures
 with [S\,{\sc ii}] fluxes published by \citet{vc92}.  From these
 we estimate an electron density of 140\,cm$^{-3}$ for both regions.
 Errors are not quoted for electron densities since published [S\,{\sc
 ii}] ratios did not give associated uncertainties.

Since the oxygen content is used as a proxy of a galaxy's metallicity, we
have derived the oxygen abundance for each region using [O\,{\sc ii}]
3727\AA\ and [O\,{\sc iii}] 5007\AA\ nebular emission lines and their
associated electron temperatures (see Table~\ref{ngc3125:properties}).  

Both regions, with log(O/H)+12 = 8.32--8.35 have an oxygen content
comparable to log(O/H)+12 = 8.37 \citep{russell90} observed in the
LMC. Our oxygen abundances are in excellent agreement with previous
studies of \citeauthor{kunth81} and \citeauthor{vc92} who obtained
log(O/H)+12 = 8.3--8.4.

\section{Massive star population}
\label{stellar}
In this section we derive the massive star content of NGC\,3125-A and
B using optical and UV techniques. 

\subsection{Estimating the number of WR stars}
 
\subsubsection{Optical}



   

For regions A and B we detect broad blue and red WR features, as
previously reported by \citet{schaerer99}. Gaussian line profiles have
been fit to the N\,{\sc iii} $\lambda 4640$/C{\sc iii} $\lambda 4650$
blend, He\,{\sc ii} $\lambda 4686$ and C\,{\sc iv} $\lambda 5808$.
Slit loss corrected emission line fluxes are presented in
Table~\ref{wr:pop}.

In region A we detect strong He\,{\sc ii} $\lambda 4686$ emission plus
N\,{\sc iii} $\lambda 4640$ -- C\,{\sc iii} $\lambda 4650$, indicative
of a predominantly late WN population.  We also detect weak N\,{\sc iv}
$\lambda 4058$, as reported by \citet{kunth81}.  This suggests a dominant
mid WN subtype \citep{smith96,pac05b}.

The presence of C\,{\sc iv} $\lambda 5808$ emission suggests that a significant
WC population is present in region A. WC stars are classified by the
C\,{\sc iii} $\lambda 5696$/C\,{\sc iv} $\lambda 5808$ line ratio \citep{smith90b}.
Since C\,{\sc iii} $\lambda 5696$ is very weak or absent, the dominant
WC population is WC4--5.

Similar WR features are observed for region B, although He\,{\sc ii}
$\lambda$ 4686 is somewhat weaker than that in region A.  As such, we
assume a dominant mid WN population whilst the red feature is again
consistent with a early-type WC population.


As a first estimate of the WR population in each region we have
derived numbers based solely on the observed He\,{\sc ii} $\lambda
4686$ and C\,{\sc iv} $\lambda 5808$ line luminosities, following the
approach of \citet{SV98}.
We have attempted to derive WR populations for A1 and A2 separately,
since UV spectroscopy relates only to A1 and A2 suffers a much higher
extinction. From Figure~\ref{figure:images}(c) we estimate that the
observed He\,{\sc ii} $\lambda$4686 flux ratio is 0.15:1 for A2:A1; whilst we
assume only A1 is responsible for the C\,{\sc iv} $\lambda$5808
emission.  This assumption is based on the higher extinction of A2
which suggests that it may be younger than A1 and would not
necessarily host a mixed WR population.

If we assume that only WN5--6 stars contribute to the He\,{\sc ii}
$\lambda$4686 line flux i.e. neglecting the WC contribution, we
estimate that N$_{\mbox{A1}}$(WN5--6)=150 based upon the $(1.8\pm1.7) \times
10^{36}$erg\,s$^{-1}$ average He\,{\sc ii} line luminosity of 15 LMC
WN5--6 stars studied by \citet{pac05b}.  For A2, assuming an internal
reddening of E$_{B-V}^{\tiny{\mbox{INT}}} \sim$ 0.5\,mag, we estimate that
N$_{\mbox{A2}}$(WN5--6)$\sim$55.  Similarly, N$_{\mbox{A1}}$(WC4)=20 based upon the average
$(3.3\pm1.7) \times 10^{36}$erg\,s$^{-1}$  C\,{\sc
iv} $\lambda 5808$ line luminosity of 7 LMC WC4 stars \citep{pac05b}.

Applying the same methodology to region B, we derive a WR population
of 75 WN and 25 WC stars.

\begin{table}
\caption{Observed WR line properties and derived populations of
  clusters in NGC\,3125.  Line fluxes (F$_{\lambda}$) are expressed in
  erg\,s$^{-1}\,\mbox{cm}^{-1}$, derived luminosities adopted a
  distance to NGC\,3125 of 11.5Mpc \citep{schaerer99} and are
  expressed in erg\,s$^{-1}$.}
\centering  
\begin{tabular}{llll}
\hline\hline
Region&A1&A2&B1\,+\,2\\
E$_{B-V}^{\tiny{\mbox{TOT}}}$& 0.24 & $\sim$0.58 & 0.21\\
\hline
F$_{4686}$&7.4$\times 10^{-15}$&$\sim$1.1$\times 10^{-15}$&4.0$\times 10^{-15}$\\
I$_{4686}$&1.7$\times 10^{-14}$&$\sim$6.8$\times 10^{-15}$&8.4$\times 10^{-15}$\\
L$_{4868}$&2.7$\times 10^{38}$&$\sim$1.0$\times 10^{38}$&1.3$\times 10^{38}$\\
\smallskip
N(WN5--6)$^{\dag}$&150&$\sim$55&75\\
F$_{5808}$&2.0$\times 10^{-15}$&---&2.8$\times 10^{-15}$\\
I$_{5808}$&3.8$\times 10^{-15}$&---&4.8$\times 10^{-15}$\\
L$_{5808}$&6.0$\times 10^{37}$&---&7.6$\times 10^{37}$\\
\smallskip
N(WC4)$^{\dag}$&20&---&25\\
F$_{1640}$&2.0$\times 10^{-15}$&---&---\\
I$_{1640}$&1.3$\times 10^{-13}$&---&---\\
L$_{1640}$&2.1$\times 10^{39}$&---&---\\
N(WN5--6)$^{\dag}$&115&---&---\\
\hline
\end{tabular}
\begin{flushleft}
\small{$^{\dag}$ Numbers derived solely from observed line luminosities based upon average line luminosities of \citet{pac05b}}
\end{flushleft}
\label{wr:pop}
\end{table}

\begin{figure*}
\begin{tabular}{cc}
\psfig{figure=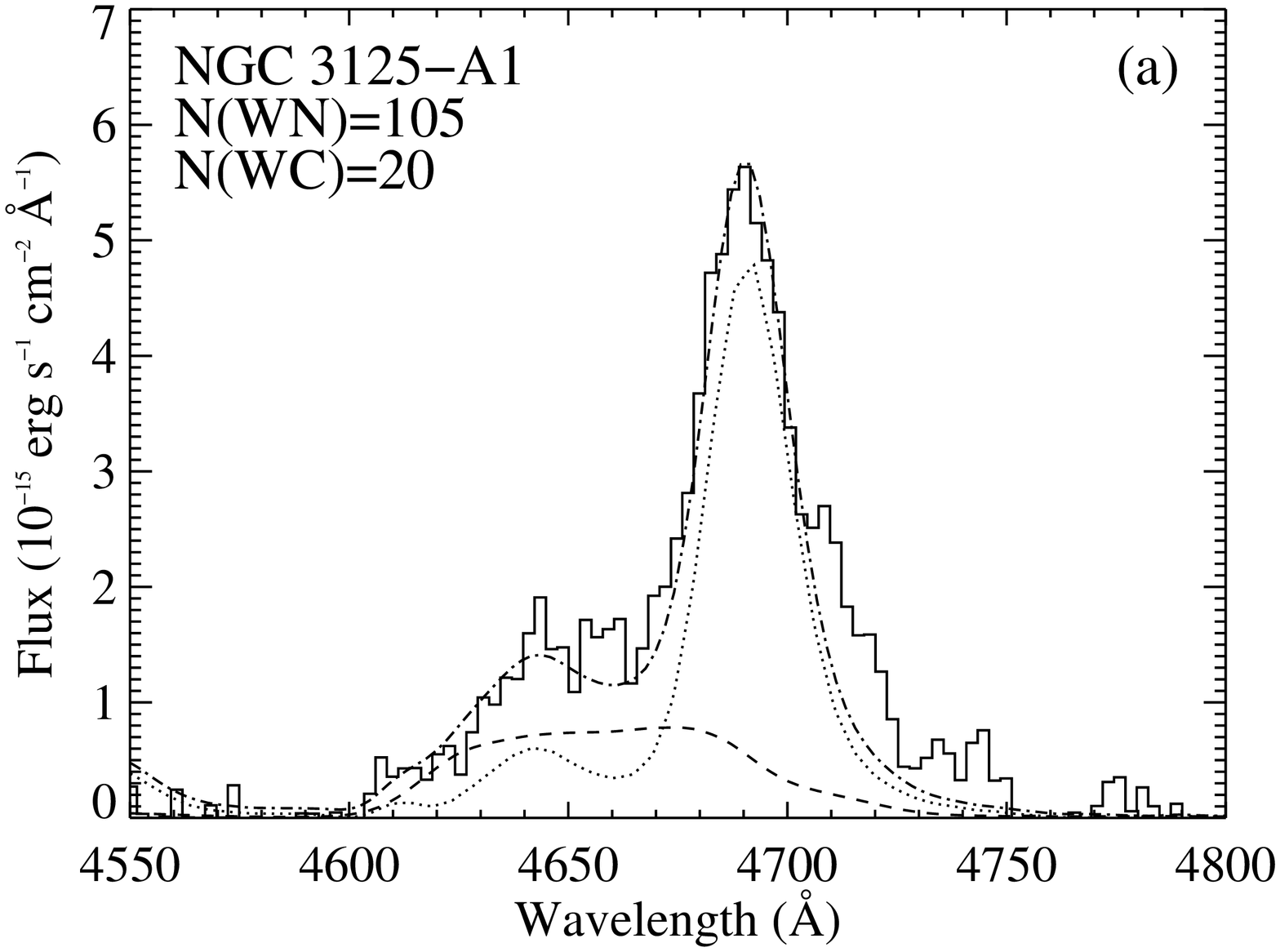,width=8cm,angle=90}&
\psfig{figure=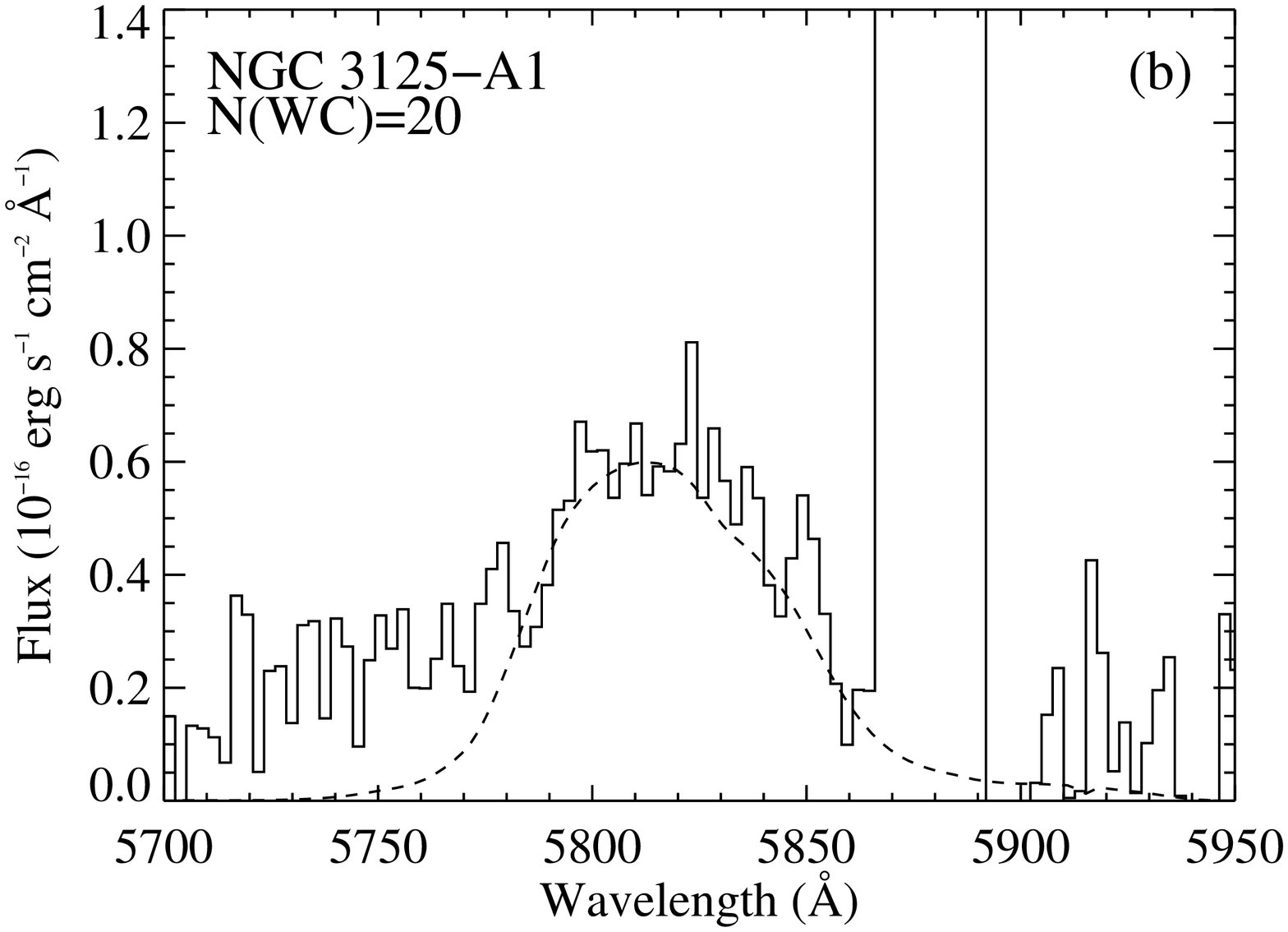,width=8cm,angle=90}\\
\psfig{figure=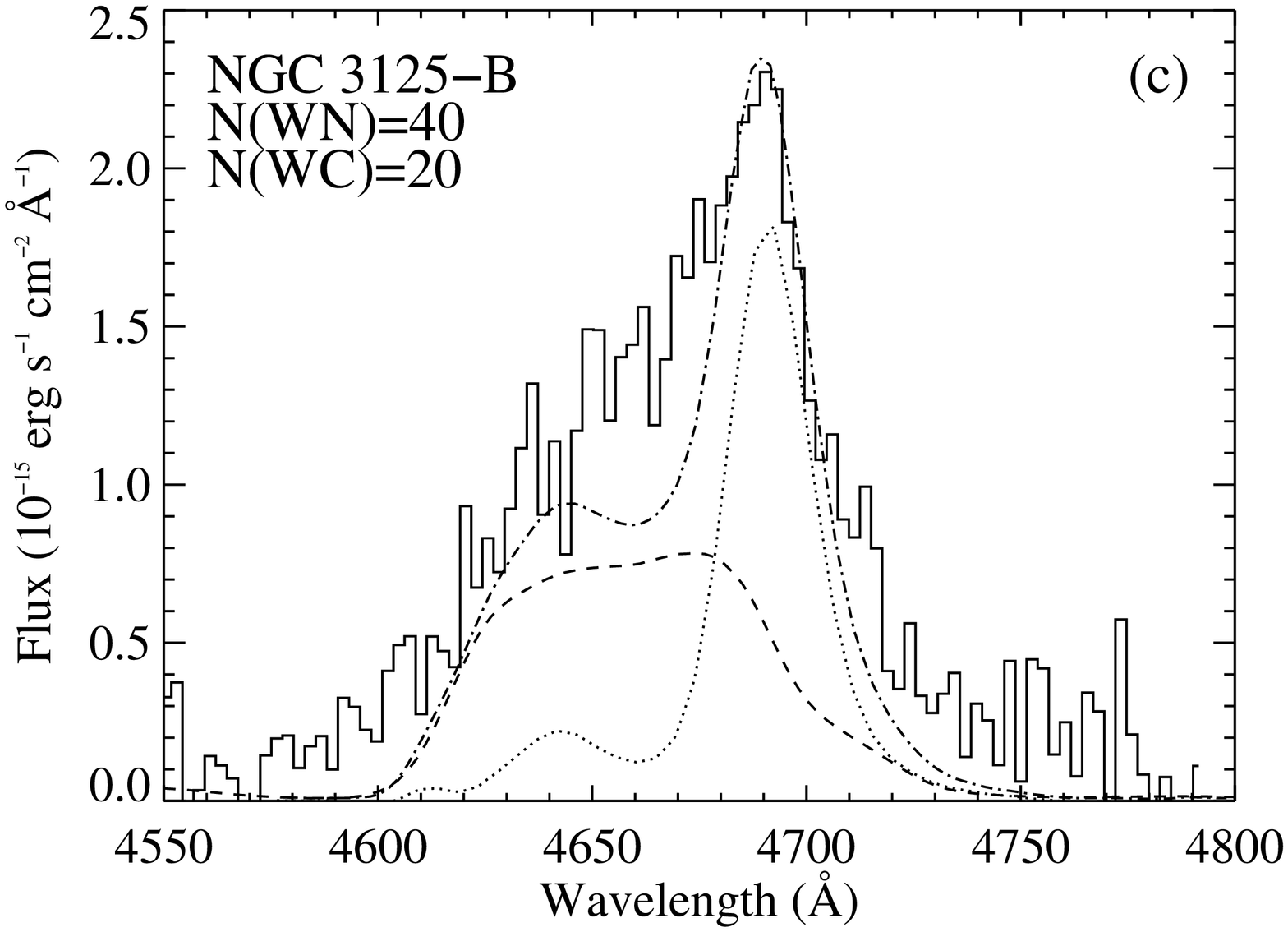,width=8cm,angle=90}&
\psfig{figure=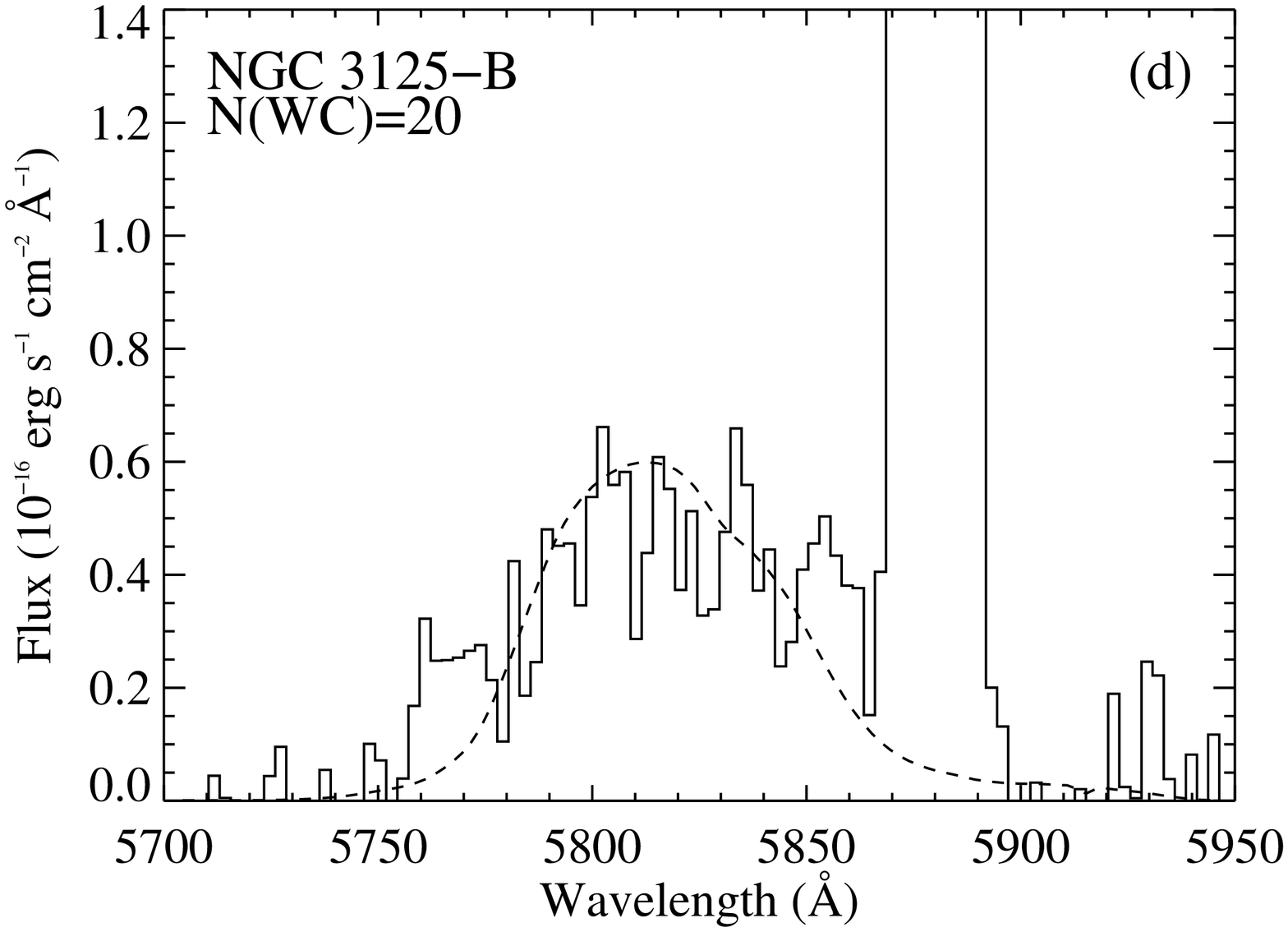,width=8cm,angle=90}\\
\end{tabular}
\caption{Dereddened (SMC extinction law), continuum subtracted,
  spectral comparison between the observed (solid) and generic
  (dashed-dotted) WR emission features for clusters A1 and B.
  Observed spectra have been velocity corrected and corrected for 25\%
  slit losses. Generic WC4 (dashed) and WN5--6 (dotted) features are
  marked.}
\label{wr:fits}
\end{figure*}

Although WN stars will typically be the primary contributor to the He\,{\sc ii}
$\lambda 4686$ feature, there will be a contribution from the WC
population if present.  Therefore, to improve upon our initial estimate of the WN
content we have estimated the WC contribution to the blue feature by
fitting generic LMC WN5--6 and WC4 spectra from \citet{pac05b} to the
observed blue and red WR bumps. Each region shall now be discussed in
turn.

In Figure~\ref{wr:fits}(b) we compare the C\,{\sc iv} $\lambda 5808$
profile of region A1 with that expected from 20 LMC-like WC4 stars
at a distance of NGC\,3125. The good agreement confirms our
initial estimate.  We have then accounted for the contribution of 20
WC4 stars to the blue WR feature, adjusted for the contribution of
A2 to the observed flux.  

In Figure~\ref{wr:fits}(a) we compare the dereddened, continuum
subtracted blue WR feature for cluster A1 with that expected for a
mixed WR population of 20 WC4 and 105 WN5--6 stars.  The composite
spectrum of our generic WR populations reproduces the observed
morphology exceptionally well, except for spectral regions where
nebular lines are expected (e.g. [Fe\,{\sc iii}] 4658, [Ar\,{\sc iv}]
4711).

For region B, it was necessary to reduce the number of WC4 stars from
25 to 20 in order to match the observed C\,{\sc iv} $\lambda 5808$
emission profile (Figure~\ref{wr:fits}(d)). This revised WC population
was used to construct the blue WR bump for region B.
Figure~\ref{wr:fits}(c) shows that the WC contribution is highly
significant, with 50\% of the He\,{\sc ii} $\lambda 4686$ and 90\% of
the N\,{\sc iii}/C\,{\sc iii} $\lambda 4640/50$ line flux originating
from WC stars.  The WN5--6 population is therefore reduced 
to 40.  WR populations derived using this method are included in
Table~\ref{table:populations}.


\subsubsection{UV}
\label{UV}

The WR population of cluster A1 has been independently estimated from
the dereddened, slit loss corrected He\,{\sc ii} $\lambda 1640$ line
luminosity.  Adopting E$_{B-V}$=0.24\,mag, as derived from
H$\alpha$:H$\beta$, we derive a dereddened line flux of
F$_{\tiny{\mbox{A1}}}$(1640)=9.4$\times 10^{-14}$ergs\,s$^{-1}$cm$^{-2}$
based upon an SMC extinction law. For completeness we have considered
various extinction laws, and conclude in Section \ref{comparison:uv}
that an SMC-extinction law provides the closest match to the complete STIS UV spectral
energy distibution.
If we neglect the WC contribution to the $\lambda$1640 line we
estimate a stellar content of 115 WN5--6 stars, based upon the average
He\,{\sc ii} $\lambda$1640 line luminosity of 1.8$\times
10^{37}$erg\,s$^{-1}$ from \citet{pac05b}.  Application of an LMC
extinction curve, which is less successful at reproducing the UV flux
distribution of A1, requires a higher extinction of E$_{B-V} =
0.33$\,mag and a WN population of $\sim$200 stars.



Following the same approach as for the optical WR features, we have
constructed a He\,{\sc ii} $\lambda 1640$ profile using generic LMC
WN5--6 and WC4 spectra.  Generic spectra are taken from \citet{pac05b}
and are based on low resolution IUE/SWP data for 10 WN5--6 stars and
medium resolution HST/FOS data for 6 WC4 stars.  


In Figure~\ref{sb99} we compare the SMC law dereddened, far-UV slit loss
corrected HST/STIS spectrum of A1, plus the spectrum around He\,{\sc
ii} $\lambda$1640 degraded to the 6\AA\ resolution of IUE/LORES --
with that for 110 WN5-6 stars from \citet{pac05b} adjusted to the
distance of NGC\,3125. Also illustrated is a synthetic Starburst99
spectrum based upon LMC/SMC template stars for the 1250--1600\AA\
region, this will be further discussed in Section~\ref{ostars}.  The
WC contribution to the He\,{\sc ii} $\lambda 1640$ flux is very minor,
reducing the number of typical WN5--6 stars from 115 to 110.  This is
in excellent agreement with the optically derived WN5--6 population
(see Table~\ref{table:populations}).

\begin{table}
\caption{WR populations for clusters within NGC\,3125 A and B derived from fitting LMC template WR spectra.}
\label{table:populations}
\begin{center}
\begin{tabular}{lllll}
\hline\hline
Region&Diagnostic&A1&A2&B1\,+\,2\\
\hline
N(WN5--6)&$\lambda$4686&105&$\sim$55&40\\
N(WN5--6)&$\lambda$1640&110&&\\
N(WC4)&$\lambda$5808&20&--&20\\
N(WR)&$\lambda$4686/$\lambda$5808&125&$\sim$55&60\\
%
\hline
\end{tabular}
\end{center}
\end{table}  


\subsection{Estimating the number of O stars}\label{ostars}

\subsubsection{The O star population of individual clusters}

\begin{figure}
\centering{
 \psfig{figure=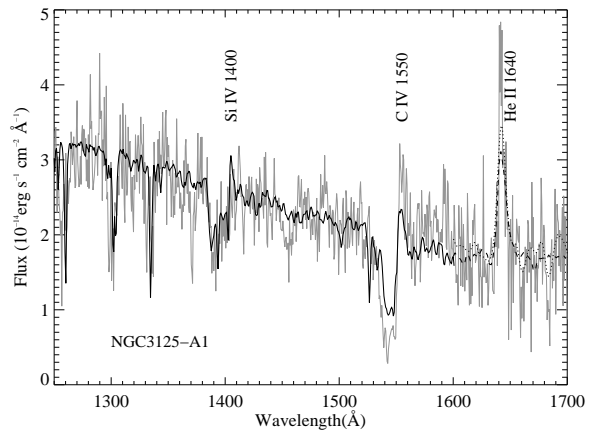,width=\columnwidth,angle=90}}
\caption{Comparison between the slit loss corrected, dereddened
(E$^{\tiny{\mbox{TOT}}}_{B-V}$=0.08(GAL) + 0.16(SMC)) HST/STIS
spectrum of A1 (faint solid) with a 2.0$\times 10^{5} \mbox{M}_{\sun}$
SB99 synthetic spectrum obtained using LMC/SMC template OB stars for
1250--1600\AA (solid), in which an age of 4Myr is indicated from the
prominent Si\,{\sc iv} wind profile.  Also shown is the STIS spectrum
around He\,{\sc ii}$\lambda$1640 degraded to 6\AA\ resolution (dotted)
together with the $\lambda$1640 emission predicted from 110 generic
LMC WN5--6 stars \citep[dashed,][]{pac05b}.}
\label{sb99}
\end{figure}

The O star content of A1 has been directly estimated by comparing the slit loss
corrected, dereddened HST/STIS spectrum of A1 to the best fitting Starburst99 model
\citep[SB99,][]{leitherer99}.  We assume a Kroupa IMF (0.1--100M$_{\sun}$) with a
turnover at 0.5M$_{\sun}$  and an exponent of 2.3 for the high mass interval.  The
Geneva high mass loss rate LMC metallicity model has been adopted. For an empirical LMC/SMC template
spectra \citep{leitherer01} we estimate a burst age of 4Myr due to the prominent Si\,{\sc
iv} $\lambda$1400 feature, see Figure~\ref{sb99}. In contrast, \citet{chandar04}
estimated 3$\pm$1\,Myr. A cluster mass of 2.0$\times 10^{5} \mbox{M}_{\sun}$ is required
to match the UV continuum, indicating an O star content of $\sim$550.  The N(WR)/N(O)
ratio for cluster A1 is therefore $\sim$0.2, explaining the large equivalent width of
He\,{\sc ii} $\lambda$1640.

If clusters A1 and A2 are coeval and using the flux ratios
described in Section~\ref{vlt:images}, we find that A2 is slightly
more massive, with M=2.2$\times 10^{5} \mbox{M}_{\sun}$.  The O star
population of A2 is estimated at $\sim$600, giving N(WR)/N(O)$\sim$0.1.  

\begin{figure*}
\begin{tabular}{cc}
 \psfig{figure=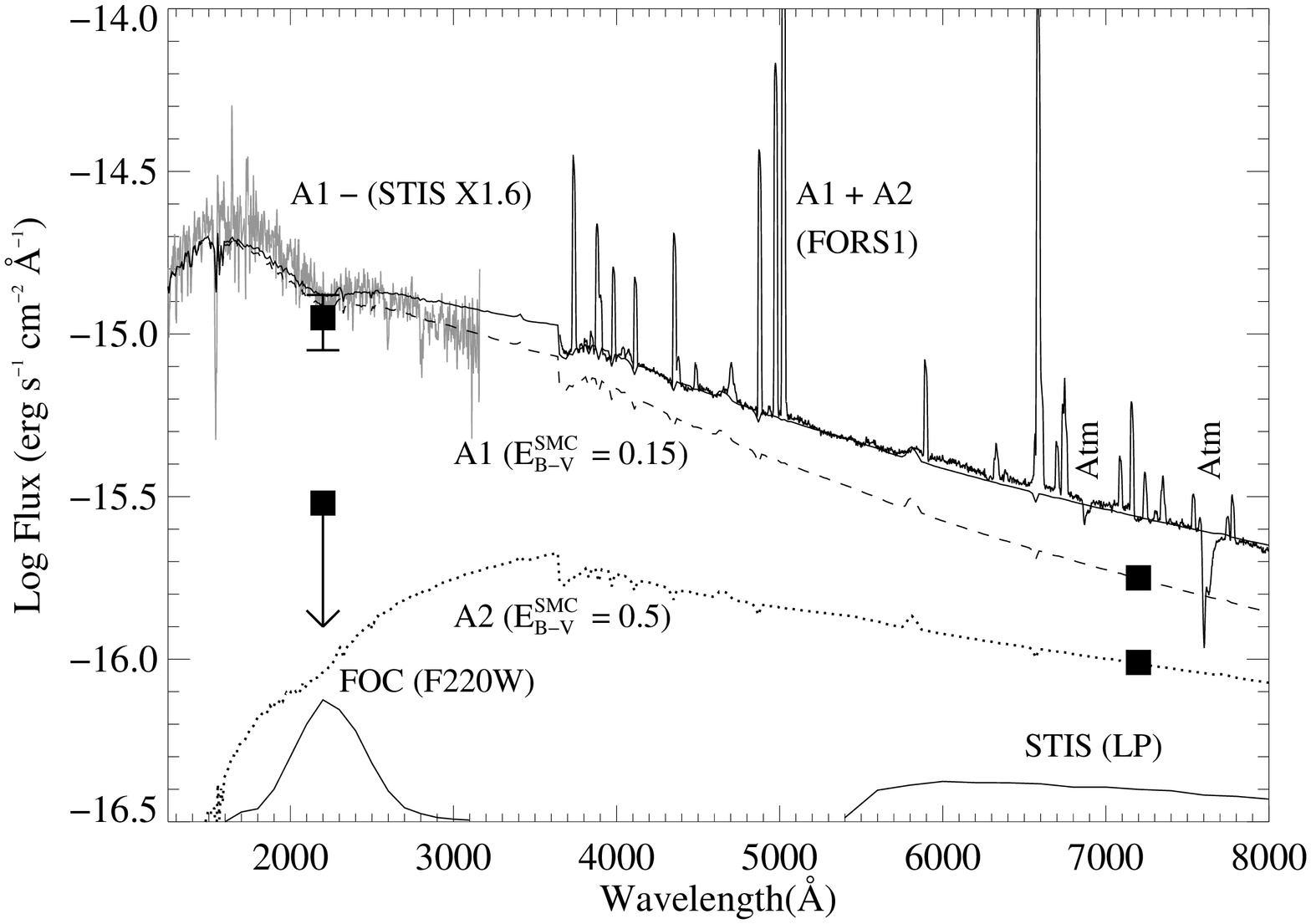,width=8cm,angle=90}&
 \psfig{figure=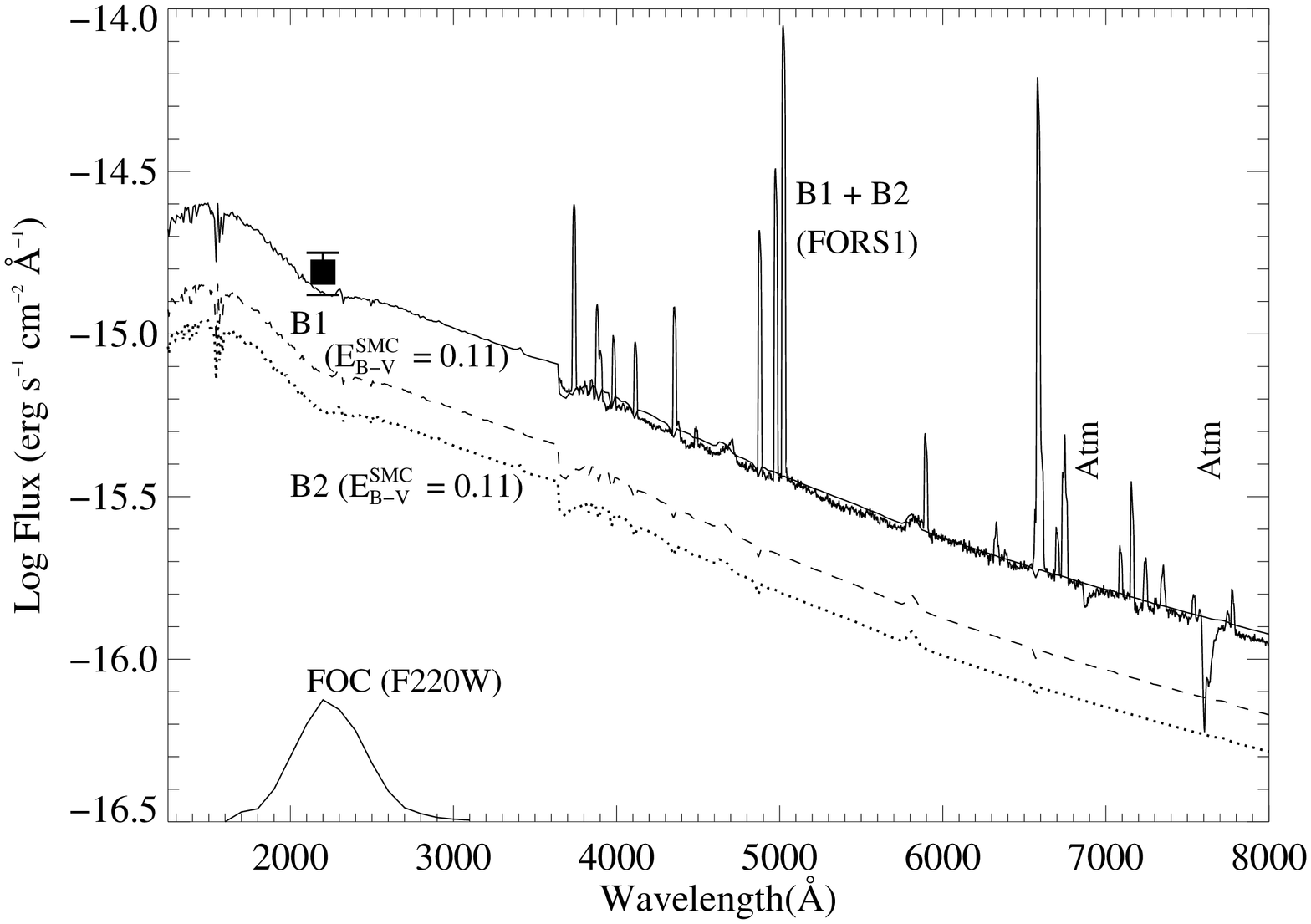,width=8cm,angle=90}\\
\end{tabular}
\caption{Left panel: Comparison between the observed VLT/FORS1 (A1+A2)
and slit loss corrected HST/STIS spectroscopy of A1. Reddened flux
distributions for A1 (M=2.0$\times 10^{5} \mbox{M}_{\sun}$ --
dashed) and A2 (M=2.2$\times 10^{5} \mbox{M}_{\sun}$ -- dotted)
are shown together with the combined, A1+A2, flux distribution
(solid). We adopt a foreground extinction of E$_{B-V}$=0.08 mag, with 
internal (SMC law) extinctions for individual  clusters indicated.  The F220W
photometry for A1 and A2 and the relative STIS/Long\_Pass flux
ratio F$_{\tiny{\mbox{A1}}}$/F$_{\tiny{\mbox{A2}}} \sim$1.9 plus the
FOC/STIS filter transmission curves are also shown.  Right panel: A
similar comparison for clusters B1 and B2 assuming identical internal
extinctions, plus the combined B1+2 FOC/F220W flux.}
\label{figure:uvsed}
\end{figure*} 
For region B, the SB99 model was scaled to match the FOC(F220W) and
VLT/FORS1 flux levels which indicates a combined mass of 1.6$\times 10^{5}
\mbox{M}_{\sun}$.  Given their relative UV fluxes, as described in
Section~\ref{hst:images}, and assuming each cluster has an identical
E$_{B-V}^{\tiny{\mbox{INT}}}$, we estimate that M(B1)=9$\times 10^{4}
\mbox{M}_{\sun}$ and M(B2)=7$\times 10^{4}\mbox{M}_{\sun}$.  SB99
models predict that such bursts should host 250 and 200 O stars,
respectively.  The O star content derived for these clusters is 450
leading to N(WR)/N(O)=0.1, as in A2.  A summary of the cluster O star content and N(WR)/N(O) ratios are presented in Table~\ref{table:O stars}. 

In Figure~\ref{figure:uvsed} we compare slit loss corrected VLT/FORS1 and
HST/STIS spectra with the combined, reddened, SB99 models for clusters within regions A and B.  Individual contributions for each cluster are indicated,
together with transmission curves for the FOC/220W and STIS/LP
filters.  
%
In order for the SB99 flux distributions of each region to match their
counterpart VLT/FORS1 spectra, we estimate that
E$_{B-V}^{\tiny{\mbox{INT}}}$(A1)=0.15 and
E$_{B-V}^{\tiny{\mbox{INT}}}$(B)=0.11, in excellent agreement with
those derived from nebular H$\alpha$/H$\beta$ and UV
spectroscopy/photometry.  Best agreement between STIS and FORS1
spectroscopy for region A1 was achieved with a slit correction factor
of 1.6 versus 1.4$\pm$0.3 obtain from HST/FOC imaging.  

\subsubsection{The O star population of the Giant H\,{\sc ii} regions A \& B}
\label{halpha}

In addition to determining the O star content of the individual
clusters from their UV/optical continuua, we have derived the number
of O stars present in each giant H\,{\sc ii} region using the net
H$\alpha$ flux measured from our VLT/FORS1 narrow-band on- and
off-H$\alpha$ images.

The observed net H$\alpha$ fluxes of regions A and B are $1.2 \
\mbox{and} \ 7.5 \times 10^{-13}$ erg\,s$^{-1}\,\mbox{cm}^{-2}$,
respectively.  For internal reddenings of 0.16 and 0.13\,mag, these
equate to H$\alpha$ luminosities of 3.2 and 2.0$\times
10^{40}$erg\,s$^{-1}$.  Measurements were made using apertures
7.0\arcsec and 6.0\arcsec in diameter for A and B, respectively.
These correspond to a physical scale of 350 and 300\,pc.  From slit
spectroscopy, H$\alpha$ luminosities are measured to be a a factor of
$\sim$3 lower.

Accounting for the WR contribution to the ionising continuum, the
number of equivalent O7V stars, $N_{O7V}$, can be expressed as
$$
N_{O7V} = \frac{Q_{O}^{Obs} - N_{WN}Q_{O}^{WN} + N_{WC}Q_{O}^{WC}}{Q_{O}^{O7V}} $$
where $Q_{O}^{Obs}$ is the observed Lyman continuum flux and
$Q_{O}^{WN}, Q_{O}^{WC}, Q_{O}^{O7V} $ are the average Lyman
continuum flux for each stellar type \citep{vc92}.

The O star population present in our regions will of course not be restricted to
the O7V spectral type, but distributed amongst the entire
spectral class.  We must therefore account for the age of the population and the
IMF when determining the total number of O stars ($N_{O}$).  As shown
by \citet{vacca94}, $N_{O}$ is related to the number of equivalent O7V
stars,
$N_{O7V}$, by
$$
N_{O}=\frac{N_{O7V}}{\eta(t)}
$$  
where $\eta(t)$ is the IMF averaged ionising Lyman
continuum luminosity for a stellar population, of an given age
normalised to one equivalent O7V star.

In recent years a number of papers \citep[i.e.][]{martins02, pac02b}
have re-calibrated the spectral~type--temperature relation for
Galactic O stars, accounting for non-LTE and line-blanketing effects.
Studies have shown that a $\sim$10\% downwards revision in the
effective temperature scale of O stars is required, such that $\log
Q_{O}$ for a typical Galactic O7V star decreases by 0.2\,dex
\citep{martins05}.  Additionally, recent studies of Magallenic Cloud O
stars indicate 2--4kK higher temperatures than their Galactic
counterparts \citep{massey05,heap06,mokiem06}.  Therefore, for this
present application we adopt $\log Q_{O7\,V}$=48.9 for LMC metallicity
stars (and would recommend 49.0 for SMC metallicities).   
For the WR contribution, we adopt $\log Q_{O}^{WC}=49.40 \ \mbox{and}\
\log Q_{O}^{WN}=49.75$, based on the average of 6 WC4 LMC stars
\citep{pac02} and 9 WN5--6 LMC stars \citep{pac96,pac98a}.

The parameter $\eta$ has been evaluated using the instantaneous
starburst models of \citet{SV98} and measured H$\beta$ equivalent
widths (W$(H\beta)$).  We measure W$(H\beta)$ to be $\sim$100\AA\ for
both regions, suggesting a burst age of $\sim$4Myr \citep[][their
Figure 7]{SV98}.  For a burst of this age and a Salpeter IMF, $\eta$
is $\sim$0.5 \citep[][their Figure 21]{SV98}.

We present $\log Q_{O}^{Obs}, N_{O7V}, N_{O}$
derived in this analysis in Table~\ref{table:O stars}.
%
%
These exceed the continuum derived cluster O star content by an order of magnitude.  These would have been reduced by a factor of three had they
been derived from H$\alpha$ slit spectroscopy.  These large difference suggest that
NGC\,3125 may host additional young massive clusters which are optically
obscured.

\begin{table}
\caption{O star populations and N(WR)/N(O) ratios
for individual clusters from UV/optical spectroscopy using Starburst99 
spectral  synthesis models, plus O star content from
giant HII regions from H$\alpha$ imaging, after correction for ionizing
fluxes from WR stars. We adopt N(LyC)=48.9 for LMC metallicity 
O7V stars, and estimate $\eta$ from W(H$\beta$) for regions A and B (see 
text)}
\label{table:O stars}
\begin{center}
\begin{tabular}{llll}
\hline\hline
Region&A1&A2&B1\,+\,2\\
\hline
\multicolumn{4}{c}{UV/optical Spectroscopy}\\
\hline
M($\times 10^{5} \mbox{M}_{\sun}$) & 2.0 & 2.2  & 1.6\\
N(O) & 550 & 600 & 450\\
N(WR)/N(O)&0.2&0.1& 0.1\\  
\hline
\multicolumn{4}{c}{H$\alpha$ Imaging}\\
\hline
$\log Q_{0}^{obs}$&\multicolumn{2}{c}{52.39}&52.19\\
N(O7V)$^{\dag}$&\multicolumn{2}{c}{2000}& 1600\\
N(O)$^{\dag}$&\multicolumn{2}{c}{4000}&3200\\
\hline
\end{tabular}
\end{center}
\end{table} 

Recall from Figure~\ref{figure:images}(d) that region A is
exceptionally bright at K$_{\mbox{s}}$.  Our combined SB99 flux
distributions for clusters A1 and A2 predict that these clusters
contribute only $\sim$30\% of the observed K$_{\mbox{s}}$ flux of
region A.  The addition of a visually obscured cluster (E$_{B-V}
\geqslant$\,1.5\,mag), spatially coincident with the optically visible
clusters may resolve the observed IR excess for region A.  To
reconcile the high number of equivalent O7V stars derived from the
H$\alpha$ image, a cluster with mass $\sim$8$\times 10^{5}
\mbox{M}_{\sun}$ and age $\sim$1--2Myr would be necessary.  Indeed,
visually obscured, young massive clusters are common in dwarf
irregular starburst galaxies such as NGC\,5253 \citep{turner00} and
He\,2--10 \citep{vacca02}.

For region B, clusters B1 and B2 are predicted to contribute
$\sim$60\% of the observed K$_{\mbox{s}}$ flux.  We do not consider it
likely that there is an additional obscured cluster since either B1 or
B2 may be somewhat older than 4Myr, such that red supergiants would
contribute to the IR excess.  In this case, the IR bright sources to
the NW of the UV/optically bright clusters would dominate the H$\alpha$
ionization from region B if each possess a mass of $\sim$2$\times
10^{5} \mbox{M}_{\sun}$ and an age of 1--2Myr.

\section{Comparison with Previous Studies}
\label{comparision}

We will now compare our derived massive stellar content for NGC\,3125
with those published in the literature.  Optically derived properties
will be compared with \citet{kunth81}, \citet{vc92} and \citet{schaerer99}
whereas UV comparisons will be made to the HST/STIS survey of
\citet{chandar04}.

\begin{table}
\caption{Comparison between observed He\,{\sc ii} 4686 and C\,{\sc iv}
5808 line fluxes in regions A and B by \citet[][KS81]{kunth81}, KS81),
\citet[][VC92]{vc92}, \citet[][S99]{schaerer99} and the present
study, including differences in interstellar reddening, with line
luminosities calculated for a common distance of 11.5Mpc.}
\begin{tabular}{@{\hspace{1mm}}l@{\hspace{1mm}}l@{\hspace{1mm}}l@{\hspace{1mm}}l@{\hspace{1mm}}l}
\hline\hline
&KS81&VC92&S99&This study\\
\hline
E$_{B-V}$(Gal+Int)& 0.4 & 0.08+0.4& 0.25+0.27  & 0.08+0.16  \\
F(4686)$_{A}$&8.2$\times 10^{-15}$&7.32$\times 10^{-15}$&8.2$\times 10^{-15}$&8.5$\times 10^{-15}$\\
L(4686)$_{A}$&1.3$\times 10^{37}$&6.1$\times 10^{38}$&7.9$\times 10^{38}$&  2.7$\times 10^{38}$      \\
F(5801)$_{A}$&--&--&3.5$\times 10^{-15}$&2.0$\times 10^{-15}$\\
\smallskip
L(5801)$_{A}$&--&--&2.2$\times 10^{38}$&6.0$\times 10^{37}$\\
E$_{B-V}$(Gal+Int)&&0.08+0.64&0.25+0.44&0.08+0.13\\
F(4686)$_{B}$&&4.3 $\times 10^{-15}$& 4.9$\times 10^{-15}$&   4.0$\times 10^{-15}$\\
L(4686)$_{B}$ &-- &8.2$\times 10^{38}$&8.4$\times 10^{38}$&1.3$\times 10^{38}$\\
F(5801)$_{B}$&--&--&6.0$\times 10^{-15}$&2.8$\times 10^{-15}$\\
L(5801)$_{B}$&--&--&2.2$\times 10^{38}$&7.6$\times 10^{37}$\\
\hline
\label{tab:comp}
\end{tabular}
\end{table}

\subsection{Optical Studies}

Table~\ref{tab:comp} compares the measured line fluxes for He\,{\sc
ii} $\lambda$4686 and C\,{\sc iv}$\lambda$5808 for regions A and B in
the present study to those from the literature, showing, in general,
very good agreement. Differences in line luminosities (for a uniform
distance), and hence WR content, relate primarily to reddening. Recall
from Section~\ref{red}, \citet{kunth81} derived reddenings from high
Balmer lines, neglecting corrections for stellar
absorptions. \citet{vc92} obtained reddenings from H$\alpha$/H$\beta$
observed during different conditions. \citet{schaerer99} applied
erroneous reddenings from the literature, including a foreground
extinction of E$_{B-V}$=0.25 rather than A$_{V}$=3.1E$_{B-V}$=0.25,
although their assumed total extinction fortuitously closely agrees
with \citeauthor{vc92}.

The major uncertainty in the derived WR
content (for our derived reddening) result from the adopted line
luminosity calibration. For WN subtypes, \citet{kunth81} adopted a
dominant mid-type WN population due to the marginal detection of
N\,{\sc iv} $\lambda$4058 emission.  We support this on the basis that
both N\,{\sc iv} $\lambda\lambda$4603-20 and N\,{\sc iii}
$\lambda\lambda$4634-41 are weak due to the contribution by C\,{\sc
iii} $\lambda$4650 from WC stars (recall Figure~\ref{wr:fits}). In
contrast, \citet{vc92} and \citet{schaerer99} adopted a dominant
late-type WN population, albeit with a similar He\,{\sc ii} line
luminosity due to their inclusion of WN6 subtypes.

For WC subtypes, both \citet{schaerer99} and the present study adopt a
dominant early-type WC population on the basis that C\,{\sc iii}
$\lambda 5696$ is weak/absent.  Both studies adopt a similar C\,{\sc
iv} $\lambda 5808$ line luminosity as \citet{pac05b} support the earlier
result of \citet{smith90} using an increased LMC sample.

Previous studies of the O star populations have been based on H$\beta$
fluxes derived from slit spectroscopy. \citet{kunth81} and
\citet{vc92} assume WR stars do not contribute to the Lyman continuum
and estimate O star populations of $\sim$2000--2400.
\citet{schaerer99} derived O star populations from H$\beta$
spectroscopic line fluxes, but with the addition of correcting for the
WR contribution and the evolution of the O star population, revising
the O star numbers for NGC3125-A and -B to 3240--6470 and 3450--6900,
respectively.  In contrast, this study has derived O star populations
using H$\alpha$ imaging for the Giant H\,{\sc ii} regions, plus
Starburst99 continuum fits to UV/optical spectroscopy for the
clusters.  For the latter, plus reduced reddenings, we estimate
N(O)$_{\mbox{A1\,+2}}$=1\,100 and N(O)$_{\mbox{B1\,+\,2}}$=450.
Differences between these and previously published results are again
primarily attributed to adopting lower reddenings.

\subsection{UV Studies}
\label{comparison:uv}

In Section~\ref{UV} we have derived a content of 110 WN5-6 stars for
NGC\,3125-A1 from the HST/STIS He\,{\sc ii} $\lambda 1640$ line flux,
whilst \citeauthor{chandar04} derived a content of 5000 late WN stars
from the same dataset (using an extraction window of 15 pixels versus
13 pixels here, and neglecting slit losses).  The origin of this major
difference is not due to the assumed intrinsic He\,{\sc ii} $\lambda
1640$ line luminosity, since we assume a higher value of 1.8$\times
10^{37}$\,erg\,s$^{-1}$ per WN star \citep{pac05b} versus 1.2$\times
10^{37}$ adopted by \citet{chandar04}. As with the optical data, we
have found that the difference between WR populations derived here and
those published by \citeauthor{chandar04} can be readily explained by
differences in the adopted internal extinction law.

The \citet{calzetti94} law was obtained using IUE large aperture
(10$\times$20\arcsec) observations of distant starbursts with a median
distance of 60\,Mpc, i.e. a physical scale of $\sim$3\,kpc.  These
observations sample a composite of stars and gas, suffering different
extinction properties. In contrast, the SMC law of \citet{bouchet85}
was obtained from IUE observations of individual stars at a distance
of $\sim$60\,kpc, i.e. a physical scale of 3\,pc. Consequently, our
HST/STIS spectroscopy (0.2$\times$0.3\arcsec) of cluster A1, sampling
a physical scale of 10$\times$15\,pc, is most naturally suited to a
low metallicity stellar extinction law since the aperture is 
dominated by stellar extinction properties rather than interstellar gas.

\begin{figure}
\centering \psfig{figure=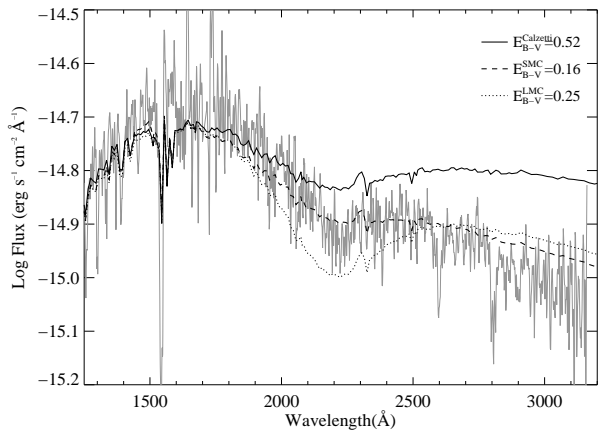,width=\columnwidth,angle=90}
\caption{
Comparison between the complete slit-loss corrected HST/STIS spectrum
of A1 (faint solid) and 4Myr, LMC-metallicity SB99 flux distributions
reddened according to (a) \citet{calzetti94} starburst extinction
law with E$_{B-V}^{\tiny{\mbox{INT}}}$=0.52 (solid); (b)
\citet{Howarth83} LMC extinction law with
E$_{B-V}^{\tiny{\mbox{INT}}}$=0.25 (dotted); (c) \citet{bouchet85} SMC
extinction law with E$_{B-V}^{\tiny{\mbox{INT}}}$=0.16 (dashed).  A
Milky Way foreground extinction of E(B-V)=0.08 has been applied in all
cases and have been normalazied at 1600\AA.
}
\label{red}
\end{figure}

In Figure~\ref{red} we compare the complete STIS spectrum of cluster A1 with various
4Myr, LMC metallicity SB99 SED (Section~\ref{ostars}) that have been reddened to
reproduce the observed 1250 -- 1600\AA\ continuum flux distribution. For the first
example the SED has been reddened using a starburst extinction law for which we
obtain E$_{B-V}^{\tiny{\mbox{INT}}}$=0.52, as derived by \citet{chandar04}. A
cluster mass of 2$\times 10^{6} \mbox{M}_{\odot}$ is needed to match the observed
1600\AA\ continuum flux.  However, the SED is highly discrepant in the near-UV, in
the sense that the predicted continuum flux is 60\% too high. Indeed, if had
selected a \citeauthor{chandar04} had selected a different wavelength interval (e.g.
1600--2000\AA) they would have derived a significantly lower reddening (e.g.
E$_{B-V}^{\tiny{\mbox{INT}}}$=0.4\,mag) for the same \citeauthor{calzetti94} law. 

In Figure~\ref{red} we also include SB99 spectral energy distributions
reddened to match the 1250--1600\AA\ slope using the SMC
\citep{bouchet85} and LMC \citep{Howarth83} extinction laws.  The
former shows good overall agreement, using
E$_{B-V}^{\tiny{\mbox{INT}}}$=0.16 and M = 2.0 $\times 10^{5} \mbox{M}_{\odot}$,
including the weak 2175\AA\ feature and gives the closest match to
the observed continuum flux in the near-UV. For the LMC law, the
far-UV slope is reproduced with E$_{B-V}^{\tiny{\mbox{INT}}}$=0.25
(3.5 $\times 10^{5} \mbox{M}_{\odot}$). In this case, however, the 2175\AA\ feature is
overestimated and the near-UV continuum flux is poorer than that for
the SMC extinction law.

Consequently, it is not possible to derive a consistent reddening for
NGC\,3125-A1 using a starburst SED with either a LMC or Calzetti
extinction law. In contrast, we obtain an excellent match to the
observed UV SED for our optically derived reddening with a starburst
spectral energy distribution and a SMC extinction law and an internal
extinction in excellent agreement with our nebular
derived H$\alpha$/H$\beta$ reddening.

\section{Conclusions}
\label{conclusions}

We have demonstrated that the WR populations of the two
regions NGC\,3125-A and B are substantially lower than previous optical
studies, and dramatically lower than previous UV studies of the bright
cluster A1 within the giant H\,{\sc ii} region A.  Indeed, previous
highly discrepant UV and optical results for NGC\,3125-A may be
reconciled using a H$\alpha$:H$\beta$ derived reddening and a SMC extinction
law for the internal extinction. We have obtained refined WR
populations in these regions by applying template spectra of typical
WN and WC stars to the blue $\lambda$4686 bump, which indicate that WC
stars, may contribute significantly to the observed $\lambda$4640
feature, commonly attributed solely to N\,{\sc iii} $\lambda$4640 from
late WN stars.

\citet{chandar04} argued that the presence of strong He\,{\sc ii}
$\lambda 1640$ emission in A1 implies an exceptional WR population
with N(WR)/N(O)$>$1.  However, evolutionary models for single stars at
LMC metallicities predict a maximum N(WR)/N(O) ratio of $\sim$0.1 for an
instantaneous burst \citep{SV98}, whilst we estimate values of
N(WR)/N(O)$\sim$0.1--0.2 for clusters within the giant H\,{\sc ii} regions
NGC\,3125-A and B.  Consequently, our results broadly reconcile the observed
massive stellar content of NGC\,3125-A and B with evolutionary
predictions for a young LMC metallicity starburst.

The primary difference between the two UV studies was the choice of
extinction law. A standard starburst extinction law is ideally suited
to spatially unresolved stellar galaxies, such as high {\it z} LBGs, but when
combined with a starburst spectral energy distribution is unable to
reproduce the UV spectrum of a resolved star cluster such as NGC\,3125-A1 or
Tol\,89-1 \citep{sidoli06}. In such cases, a SMC (or LMC)
reddening law is necessary to correct the complete UV spectrum for
extinction. Consequently, He\,{\sc ii} $\lambda 1640$ results for
other nearby WR clusters studied by \citeauthor[][]{chandar04}
should be treated with caution for the present.

We have also demonstrated that the O star content derived from
H$\alpha$ narrow band imaging is substantially higher than
that estimated from continuum flux techniques.  On the basis of
near-IR imaging of NGC\,3125 we propose that the discrepancy for
region A can be resolved by the presence of an additional cluster
which is optically obscured.  Such clusters appear to be common in
LMC-like metallicity starburst galaxies such as NGC\,5253 and He2--10
\citep{turner00, vacca02}.  For region B, the O star content may be
resolved by the identification of two additional knots to the NW of
region B which are bright at near-IR wavelengths.

The present study emphasises the need for the highest possible 
spatial resolution. We have attempted to evaluate the properties of the 
two optically visible clusters within region A based upon a pair 
of broadband  ultraviolet (FOC) and far-red (STIS) flux ratios, plus the 
properties of the clusters within region B from a single, composite 
ultraviolet  measurement. Refined properties for clusters within NGC~3125 
require high  spatial resolution datasets spanning the ultraviolet to 
near-infrared, as recently undertaken by program GO~10400 (R.~Chandar, 
P.I.)  using the HST Advanced  Camera for Surveys (ACS) and Near infrared 
Camera and Multi-Object Spectrometer (NICMOS).

\section*{Acknowledgements} 
We wish to thank Bill Vacca for providing us with the ``red'' CTIO
spectrum of NGC\,3125-A and for various communications which helped
with the analysis. We also appreciate suggestions made by an anonymous
referee which helped improve the manuscript.  
Some of the data presented in this paper were based on observations
made with the NASA/ESA Hubble Space Telescope, obtained from the data
archive at the Space Telescope Institute. STScI is operated by the
association of Universities for Research in Astronomy, Inc. under the
NASA contract NAS 5-26555.
LJH acknowledges financial support from PPARC, PAC acknowledges
financial support from the Royal Society.

\end{document}